\documentclass[aps,pre,reprint,notitlepage,longbibliography,showpacs]{revtex4-1}
\usepackage{times}
\usepackage{amsmath}
\usepackage{float}
\usepackage{url}
\usepackage{graphicx}

\newcommand{\braket}[1]{\ensuremath{\left\langle #1 \right\rangle}}

\usepackage[
pdfauthor={Sadjadi,Thorpe},
pdftitle={Ring correlations in random networks},
pdfstartview=XYZ,
bookmarks=true,
colorlinks=true,
linkcolor=red,
urlcolor=blue,
citecolor=blue,
pdftex,
linktocpage=true, % makes the page number as hyperlink in table of content
hyperindex=true
]{hyperref}

\begin{document}

\title{Ring correlations in random networks}

\author{Mahdi Sadjadi}
\email[Email:]{mahdisadjadi@asu.edu}
\homepage[Web:]{http://mahdisadjadi.com}
\affiliation{Department of Physics, Arizona State University, Tempe, AZ
85287-1504, USA}

\author{M. F. Thorpe}
\email[Email:]{mft@asu.edu}
\homepage[Web:]{http://thorpe2.la.asu.edu/thorpe}
\affiliation{Department of Physics, Arizona State University, Tempe,
AZ 85287-1504, USA \\ and Rudolf Peierls Centre for Theoretical Physics,
University of Oxford, 1
Keble Rd, Oxford OX1 3NP, England}

\begin{abstract}
We examine the correlations between rings in random network glasses in two
dimensions as a function of their separation. Initially, we use the topological
separation (measured by the number of intervening rings), but this leads to
pseudo-long-range correlations due to a lack of topological charge neutrality
in the shells surrounding a central ring. This effect is associated with the
non-circular nature of the shells. It is, therefore, necessary to use the
geometrical distance between ring centers. Hence we find a generalization of
the Aboav-Weaire law out to larger distances, with the correlations between
rings decaying away when two rings are more than about 3 rings apart.
\end{abstract}

\maketitle

\section[Introduction]{Introduction}
The structure of network glasses is often described by continuous random network (CNR)
model. In this model, building units form a random network where short-range order is
preserved similar to that in crystals but translational long-range order is absent
due mainly to distorted bond angles \cite{rosenhain1927structure,zachariasen1932atomic,
wright2013eighty}. Such structures have been generally studied by models
\cite{gupta1990topologically} and diffraction experiments \cite{warren1992x}
which have provided invaluable information on short-range and medium-range
order, mostly in the form of pair distribution functions (PDFs)
 \cite{wright1994neutron,elliott1991medium,thorpe2012properties,
treacy2005fluctuation}.

One challenge in using diffraction data is that this only provides average properties
such that the structure cannot be reconstructed uniquely. Meanwhile,
Scanning Probe Microscopy (SPM) and Electron Microscopy (EM) techniques have radically
shortened the resolution limit and recently true atomic resolution images of silica
bilayers and other two-dimensional (2D) amorphous surfaces have become
available \cite{lichtenstein2012atomic,huang2012direct}. However, high
resolution imaging of bulk amorphous materials remains elusive \cite{burgler1999atomic}.
These new results on 2D glasses have opened up numerous opportunities to
study the structure of glasses using actual atomic
coordinates. Recent work on 2D glasses includes modeling of silica
bilayers \cite{wilson2013modeling,wilson2015modelling}, ring
distribution \cite{kumar2014}, medium-range order \cite{buchner2016building},
suitable boundary conditions to recover missing
constraints in the surface \cite{theran2015anchored} and the refinement of experimental
samples \cite{sadjadi17refining}. Rigidity theory has also uncovered a
connection between 2D glasses and jammed disk
packings \cite{thorpe1983continuous,ellenbroek2015rigidity}.

The remarkable images of vitreous bilayer silica (SiO$_{2}$) unveil a ring
structure which is the characteristic of covalent glasses. But similar underlying
structure also can be found in various amorphous materials such as amorphous graphene
\cite{kotakoski2011point,kumar2012amorphous,buchner2014topological}. In fact,
these atomic materials are members of a larger class of materials
(many with larger length scales) collectively
known as \textit{cellular networks}. Examples are foams and grains \cite{sadoc%
2013foams}, biological tissues \cite{mombach1993two}, metallurgical aggregates,
geographical structures, crack networks \cite{korneta1998topological}, ecological
territories, Voronoi tessellations \cite{weaire1984soap, stavans1993evolution}
and even the universe at large scale \cite{aragon2014universe} and
fractals \cite{schliecker2001scaling}. Given wide range of length scales,
formation mechanisms and physical properties, cellular networks have been subject
of many studies \cite{gibson1999cellular,schliecker2002structure}. Despite the
topological resemblance between 2D amorphous systems and other cellular networks,
one should note that these materials are microscopic systems with a very
different nature of bonds and forces and hence they can shed light on new
properties of cellular networks, in particular those related to geometry.

\begin{figure}[b!]
  \includegraphics[width=4.0cm,height=4.0cm]{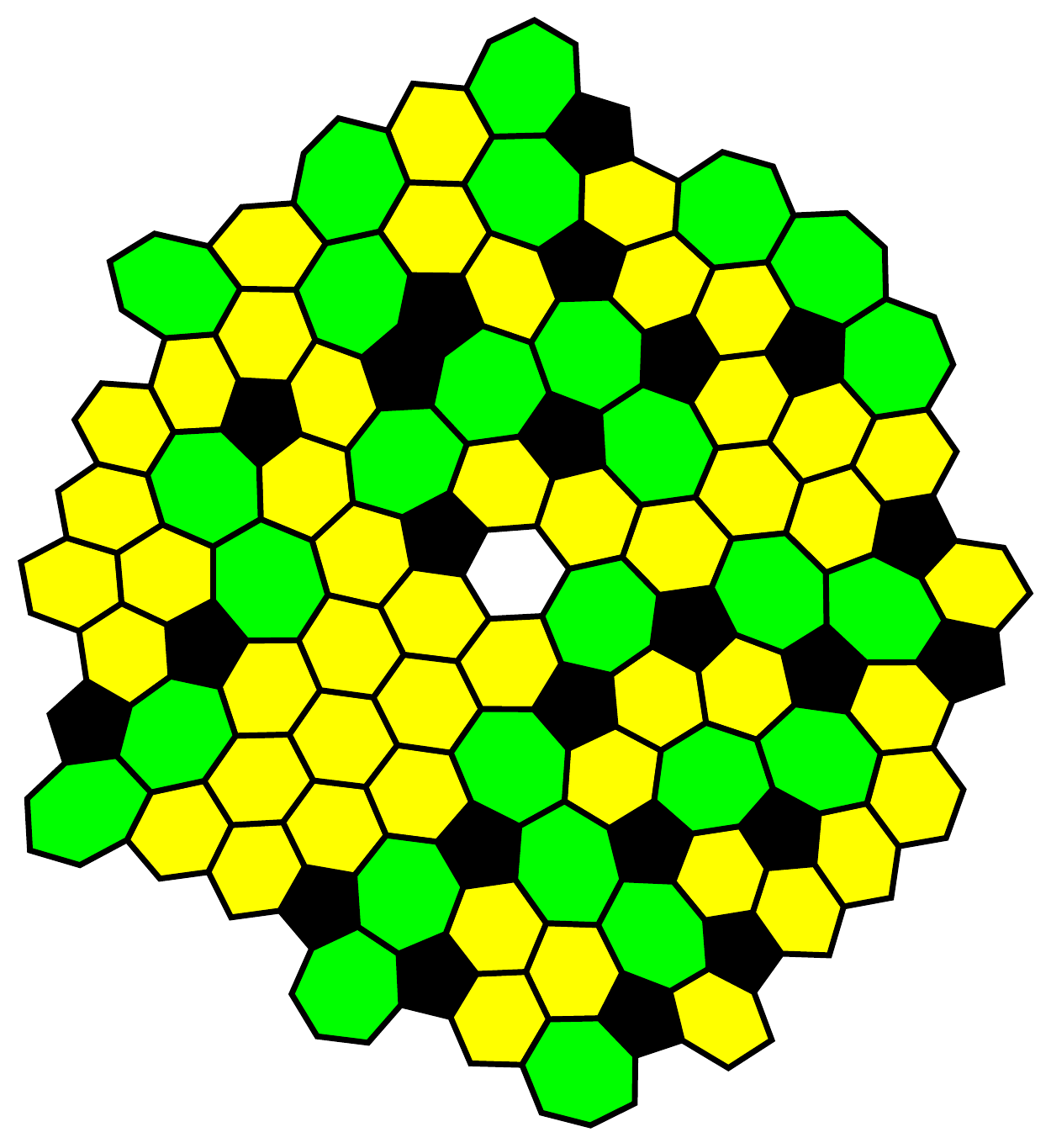}\quad
  \caption{\label{fig:randnet} A piece of a two dimensional cellular network generated by
  bond-switching algorithm from a honeycomb lattice. Rings are colored based
  on their size. On the bottom left corner, a group of six-fold rings can be
  seen which also happens in experimental samples and is a feature of
  amorphous materials, due to statistical correlations. A central six fold ring
  has been left uncolored and shells of rings will be found around this. Any
  ring can be used as a central location.}
 \end{figure}

These glassy networks are almost entirely $3-$coordinated networks, i.e.,
each vertex is connected to three other vertices through edges which form the
boundary of polygonal rings (Fig. \ref{fig:randnet}). In the case of
amorphous graphene - vertices represent carbon atoms. In silica bilayer, rings
are formed by connecting silicon atoms while intervening oxygen atoms
 are omitted.

These glassy networks, to some extent, are random and their study requires a
statistical approach but experimental samples of amorphous materials
are relatively small \cite{klemm2016preparation}. Additionally, the small size of
many samples does not permit the study of ring correlations at larger distances
with good statistics. In this work, we employ large computer
models to study correlations among the rings.
In the literature, the focus has been on the correlation among adjacent
rings where well-known Aboav-Weaire's law captures the tendency of smaller
and larger rings to be adjacent. This paper studies various correlation
functions out to large topological and geometrical distances and generalizes the
Aboav-Weaire's law.

\begin{figure}[t!]
  \includegraphics[scale=0.27]{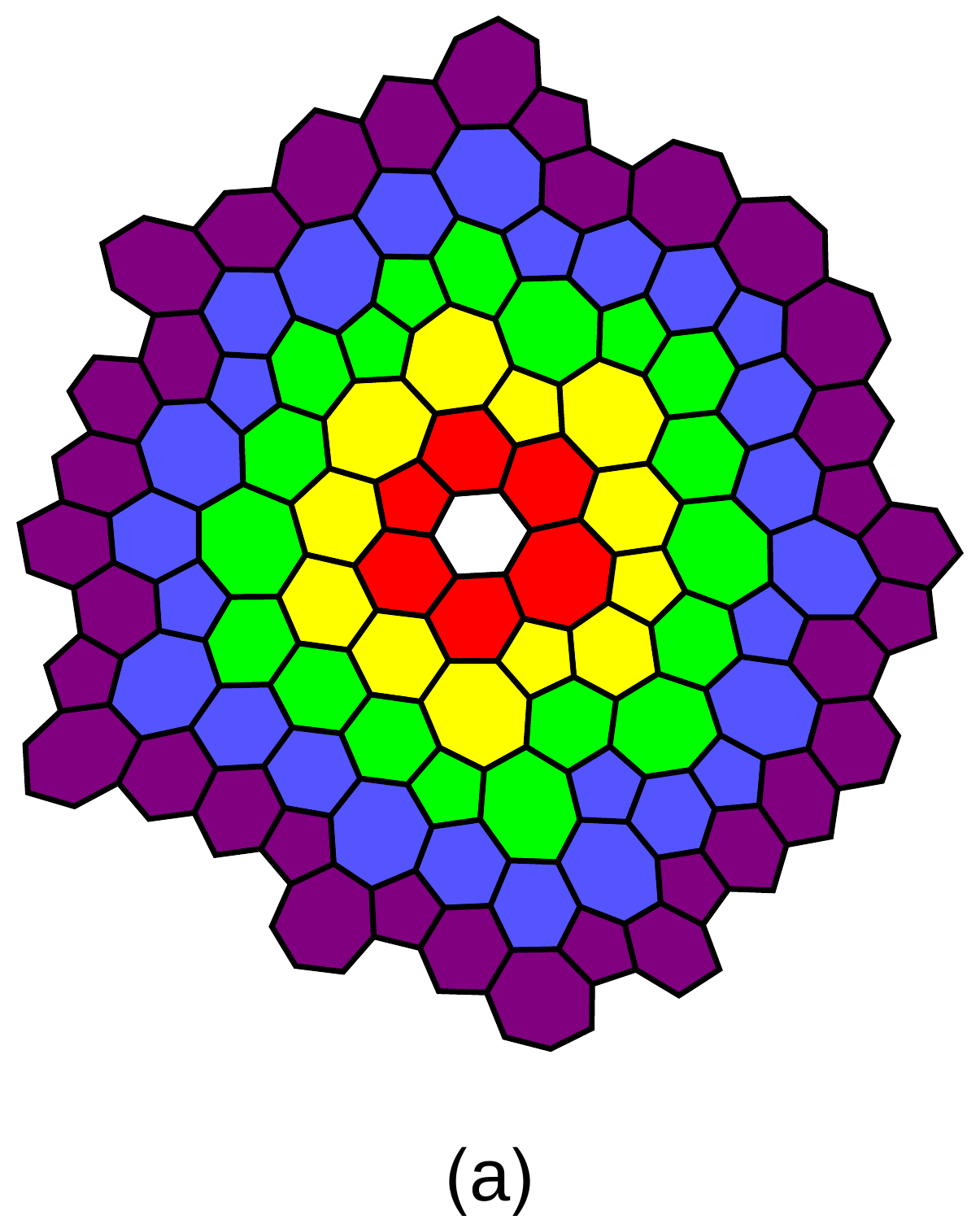}\quad
  \includegraphics[scale=0.27]{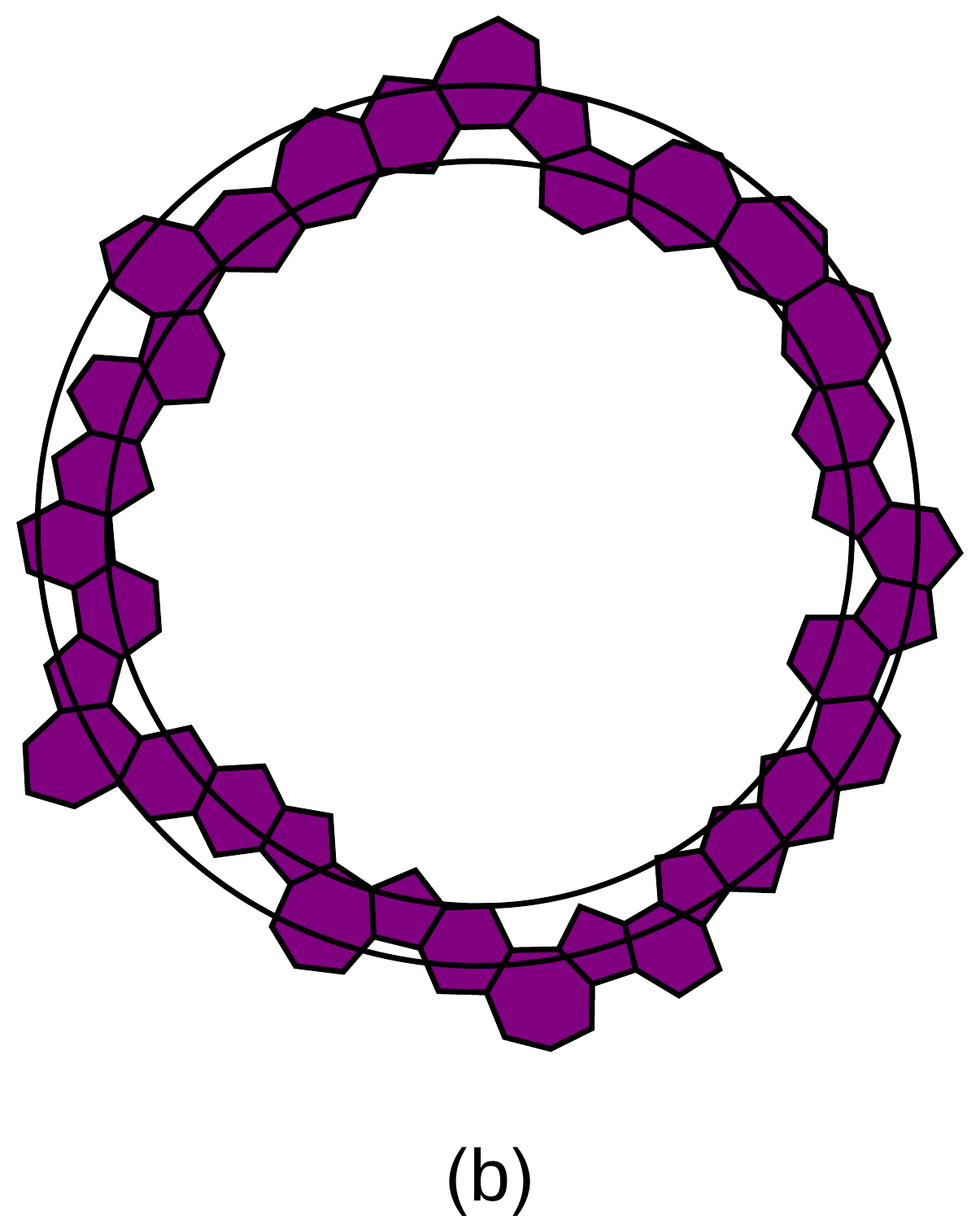}
  \caption{\label{fig:dis} (a): Partitioning of the random network in Fig.
  \ref{fig:randnet} into
  topological shells. The shells grow roughly in circular shapes.
  This piece also has a triplet inclusion in the forth (blue) shell where a
  5-ring is isolated from the fifth (purple) shell. (b) Although shells are
  roughly circular, no circle can sweep all rings within a single shell; hence
  ring distributions with topological and geometrical definitions are different.}
 \end{figure}

\section{Shell analysis and correlations}\label{sec:framework}
We define an $n-$ring as a ring with $n$ adjacent rings. The ring distribution
of a network with a total of $N$ rings is characterized by $p(n)$, the fraction of
$n-$rings, its mean $\braket{n} = \sum_{n} n p(n)$, and the second moment about
the center $\mu = \braket{n^2} - \braket{n}^2$. According to Euler's theorem,
mean ring size for a network with periodic boundary conditions (PBCs) is exactly
$\braket{n} = 6$ \footnote{The mean ring size in the finite experimental
samples is slightly less since the surface sites are
under-coordinated. Although for sufficiently large
systems, boundary effects are negligible.}. The ensemble
average of a quantity $x$ is defined as $\braket{x} = \sum_{n} p(n) x$.
To overcome the finite size effect in the experimental samples,
we use computer-generated models under PBCs with $\sim$100000 vertices
($\sim$50000 rings) generated from an initially honeycomb lattice using
bond-switching algorithm. Here,
a bond between two nearest neighbor sites is selected and replaced by a dual bond
at right angle and local topology is reconstructed to maintain the three-fold
coordination everywhere~\cite{wooten1985computer,stone1986theoretical}. Although,
experimental samples contain rings with size $4$ to $9$, but fraction of rings
with sizes other than $5$ to $7$ are statistically quite rare \cite{kumar2014}.
We studied two networks one with only $5$ to $7$ rings and one with $5$ to $8$
but no essential difference was observed. Therefore we report results
of the network with $5$ to $8-$fold rings with the following ring
distribution: $p(5) = 0.262$, $p(6) = 0.494$,
$p(7) = 0.227$, $p(8) = 0.0172$ and $\mu = 0.558$.
Nevertheless, the measures of this paper are general and can be applied
to all glassy and cellular networks.

The correlation among rings is usually defined over
a \textit{topological distance} $t$. The topological distance between two rings
is defined as the minimum number of bonds should be traversed to connect two
rings. This distance is the equivalent of distance of two nodes in the dual graph
(when each ring is represented by a node) of Fig. \ref{fig:randnet}.
The distance of a ring from itself is zero ($t=0$). All rings
which have one common side with a given central ring are located at $t=1$
(first shell). Adjacent rings to the first shell, excluding the central ring,
are at $t=2$ (second shell). This process can be continued to find
shells at any topological distance similar to Fig. \ref{fig:dis}. A ring at shell
$t$ is adjacent to at least one ring at shell $t-1$ and usually adjacent to
at least one ring in shell $t+1$, otherwise this ring is
trapped and forms a \textit{triplet inclusion} (Fig. \ref{fig:dis}).
This definition naturally divides/partitions the network into concentric shells
around any given ring. Therefore, all properties of the network are studied as a
function of the topological distance and the size of the central ring
\cite{fortes1993average,mason2015geometric}, as first pointed out by Aste et
al \cite{aste1996statistical, aste1996one}.

A shell at distance $t$ from an $n-$ring is characterized by
three numbers: number of $n'$-rings $N_t(n,n')$; total number of rings (shell
size) $K_t(n)$, and total number of sides (edges) $M_t(n)$. These quantities
are related as follows:
\begin{align}
 K_t(n) &= \sum_{n'} N_t(n,n'), \label{eq:shellsize} \\
 M_t(n) &= \sum_{n'} n' N_t(n,n'). \label{eq:edgenumber}
\end{align}
Since these equations are linear, they are also valid for the averaged values over
all $n-$rings. More importantly, note that $N_t(n,n')$ is not symmetric in respect
to $n$ and $n'$. This reflects the fact that local order of the rings
is strongly dependent on the size of the central ring.
Specially,
$N_t(n,n')$ should not be confused by the number of
$n-n'$ pairs at topological distance $t$:
\begin{equation}\label{eq:symfun}
 N p(n) N_t(n,n') = N p(n') N_t(n',n),
\end{equation}
which by definition is symmetric.
This symmetry can relate the ensemble average
of the number of sides (Eq. \ref{eq:edgenumber}) to the ensemble average of shell
size (Eq. \ref{eq:shellsize}) at any topological distance:
\begin{align}\label{eq:gwsr}
\braket{M_t} &= \sum_{n} p(n)M_t(n)
               = \sum_{n}\sum_{n'}p(n) n' N_t(n,n')  \nonumber \\
              &= \sum_{n'} n' p(n') K_t(n')
               = \braket{n K_t}.
\end{align}
This relation is the generalized Weaire sum rule which was originally
proposed for the first shell where it
takes the form $\braket{M_1}=\braket{n^2} = \braket{n}^2 + \mu$
\cite{weaire1974some,lambert1983order}. Note that the first shell is the
only shell that $K$ is exactly determined $[ \,K_1(n)=n] \,$ but Eq. \ref{eq:gwsr}
surprisingly encapsulates all the statistical variation in the local ring
distribution in a simple form.

\begin{figure}[t]
  \centering
  \includegraphics[width=5.8cm,height=5.8cm]{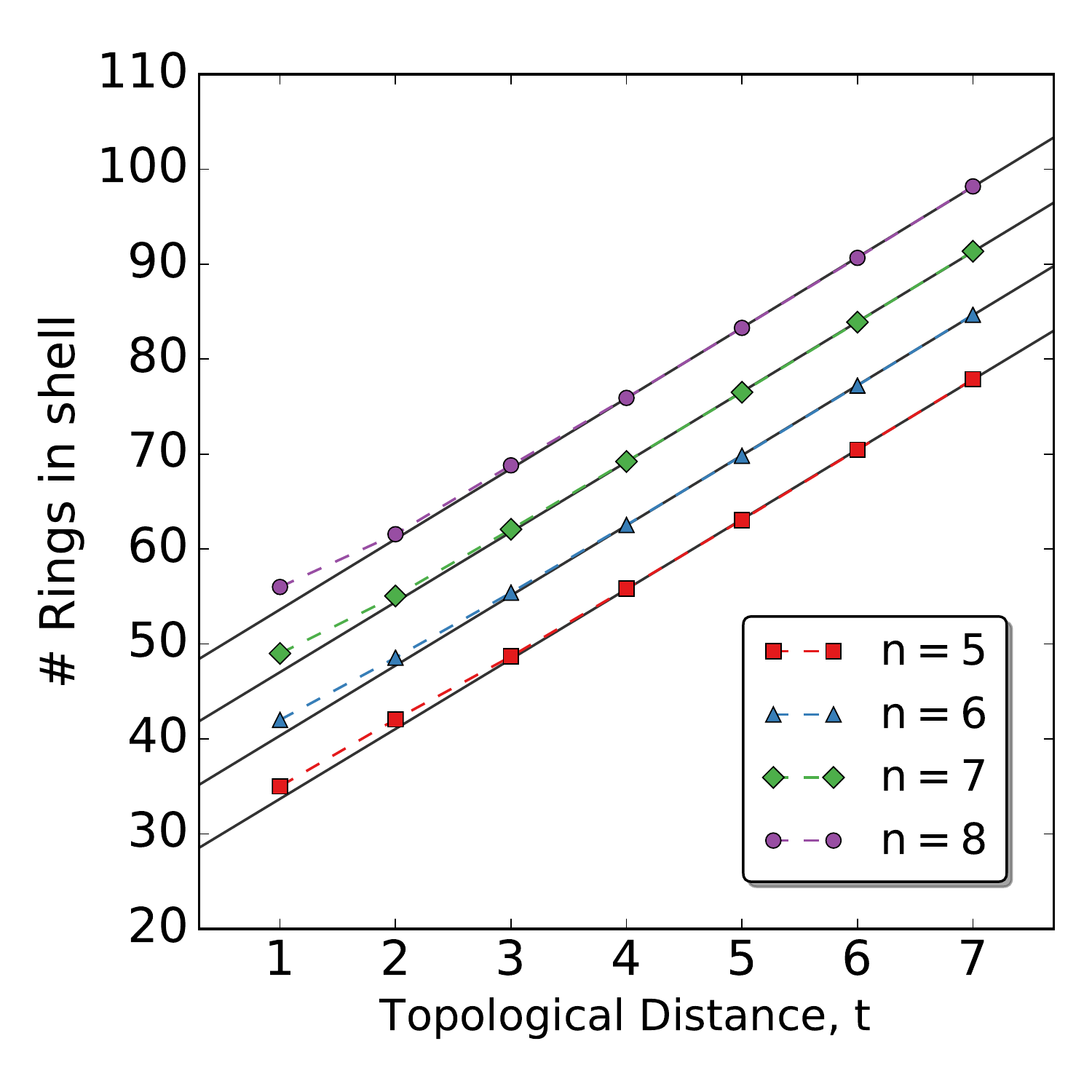}
  \caption{
  \label{fig:shellsize}
  Dependence of the number of rings $K_t(n)$ on topological distance $t$ and
  size of the central ring $n$. $K_t(n)$ grows linearly for $t \ge 4$. Solid lines
  are fitted lines to the last three points. Points are offset for clarity with
  $6n$.}
\end{figure}

The space-filling nature of rings in the network requires that
$K_t(n)$ scales linearly with $t$ in the absence of correlation. This means that the
growth rate of the shell size is a constant number independent of the size of
the central ring. Although, geometrical constraints on the polygonal tiling of
the plane does not allow a complete independence from the central ring simply
because shell closure around a larger ring requires more rings. As a result,
the intercept of $K_t(n)$ remains a function of $n$. Therefore
we expect that:
\begin{equation}\label{eq:linform}
 K_t(n) = A t + B(n),
\end{equation}
for $t \ge \xi$, where $\xi$ is the ring correlation length.
In a hexagonal lattice, the growth rate $A$ is 6 but as Fig. \ref{fig:dis}
shows, in a random network, shells grow roughly in circular form and simple
geometrical arguments predict that the growth rate should be $2\pi$.
However, because rings meet each other at random orientations and the shell
surface is rough, the actual growth rate is usually greater than $2\pi$ and
$A$ can be a measure of this roughness \cite{oguey2001roughness}.
Figure \ref{fig:shellsize} shows the number of the rings in the shells around
different central rings. The linear behavior of the shell size is observed in
various systems and is present in 2D glass, as expected. However, in 2D glasses
$A=7.31\pm0.1$ which is much smaller compared to the reported
values for Voronoi tessellation ($11.0\pm0.2$) and soap ($9.45\pm0.1$)
\cite{aste1996statistical}, probably due to the bond bending interactions
which result in the high symmetry (close to the maximum area forgiven edge
lengths) of the rings in the 2D glass \cite{kumar2014}.

Another useful quantity is the topological charge of an $n$-ring defined as
$6-n$. Since the mean ring size in the network is $6$, equivalently total charge
of the network is zero. However a piece of the network can contain any
amount of charge depending on the local ring distribution. Hence, topological
charge is a useful quantity that monitors the local deviation from the bulk
properties. In particular, the topological charge of a shell $q_t(n)$ can
be defined as the sum of the charge of its rings:
\begin{equation}\label{eq:shellcharge}
 q_t(n) = \sum_{n'} (\braket{n}-n') N_t(n,n') = \braket{n} K_t(n) - M_t(n).
\end{equation}

\begin{figure}[t]
  \centering
  \includegraphics[width=5.8cm,height=5.8cm]{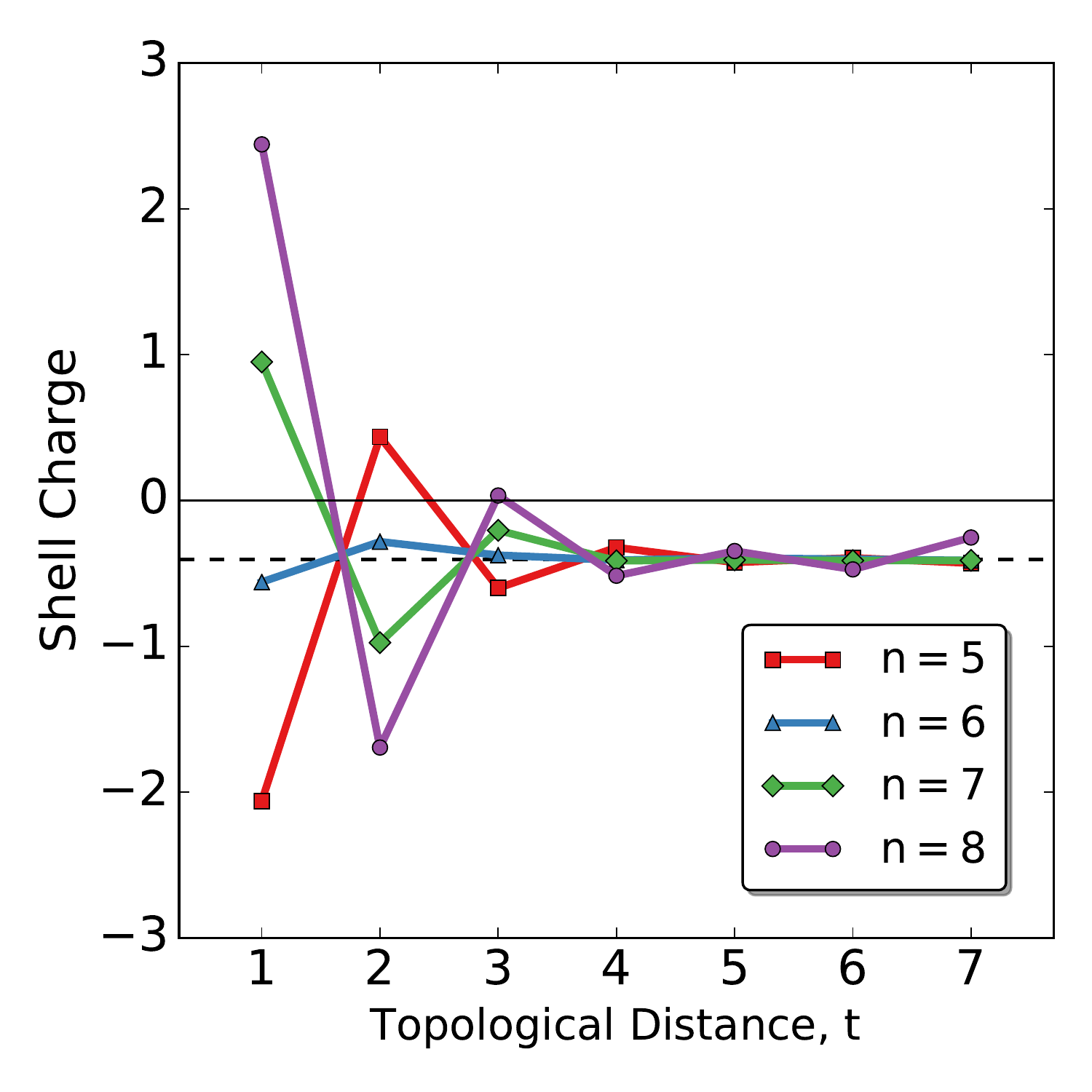}
  \caption{\label{fig:charge}
  Shell charge $q_t(n)$ vs. topological distance $t$. The shell charge
  settles to a constant non-zero number for $t~\ge~4$. The dashed line shows the
  asymptotic offset $-0.4$.}
\end{figure}

From short- and medium-range order, it is expected that rings around a given
ring are distributed such that the charge of the central ring is screened by
the charge of the neighboring shells and for $t>\xi$, the ring distribution is
similar to the bulk (charge per
shell is zero). But as Eq. \ref{eq:linform} shows, the shell size is a function
of $n$ for any distance and therefore rings are counted with different weights
in calculating the charge per shell. In fact, Eqs. \ref{eq:gwsr},
\ref{eq:linform} and \ref{eq:shellcharge} readily
yield an asymptotic value for the shell charge for $t>\xi$:
\begin{equation}\label{eq:asymcharge}
\braket{q_t} = \braket{(\braket{n}-n)K_t}
\approx p(5)B(5)-p(7) B(7),
\end{equation}
which is exact for a network with $n=5,6,7$ and approximately correct
as long as fraction of the other rings is negligible. Therefore $\braket{nK_t}$
does not factorize and statistically, there is a tendency to have
larger rings in a shell $[ \,\braket{q_t}<0$ since $B(7)>B(5)] \,$ .

The results of calculating the charge per shell is shown in Fig.~\ref{fig:charge}.
For $t=1$, the total shell charge has an opposite sign to the charge of the
central ring to screen the charge but for $t>1$ screening does not happen and the
charge per shell reaches a non-zero constant value, conjectured in Eq. \ref{eq:asymcharge}.
It is interesting to note that although the charge of $5$- and $7$-rings have
the same magnitude, the strength of screening for these two is considerably
different in the first shell. This shows that geometry has a strong effect on the
ring distribution. Note that hexagons have short-range correlations ($\xi=1$)
but other rings are correlated up to $\xi=3$ (medium-range correlation)
with different strengths.

Topological charge gives a rather complete picture of correlations in the
shell structure, but the most studied measure of correlations in the literature
is the mean ring size in the first shell around a central ring, through the
well-known Aboav-Weaire law that a ring with large size tends to have smaller
rings in its neighborhood and vice versa \cite{chiu1995aboav,
mason2012geometric}. Mathematically, the mean ring size $m_1(n)$ around a ring with
$n$ neighbors can be written (to a very good approximation)
as \cite{aboav1970arrangement,weaire1974some}:
\begin{equation}\label{eq:awl}
 n m_1(n) =  \braket{n}^2 + \mu + \braket{n} (1-\alpha)(n-\braket{n}),
\end{equation}
where $\alpha$ is a fitting parameter which depends on the specific network.
Usually a network is characterized by
$(\mu,\alpha)$. The meaning of $\alpha$ is not clear but it has been argued that
it is a metrical quantity \cite{aboav1984arrangement} or the average excess
curvature \cite{mason2012geometric} but these definitions only work in special
cases. In our network, $\alpha\approx0.23$ which is somewhat smaller than values
extracted from experiments \cite{kumar2014} showing computer generated
models still need further refinement.

\begin{figure}[t]
  \centering
  \includegraphics[width=5.8cm,height=5.8cm]{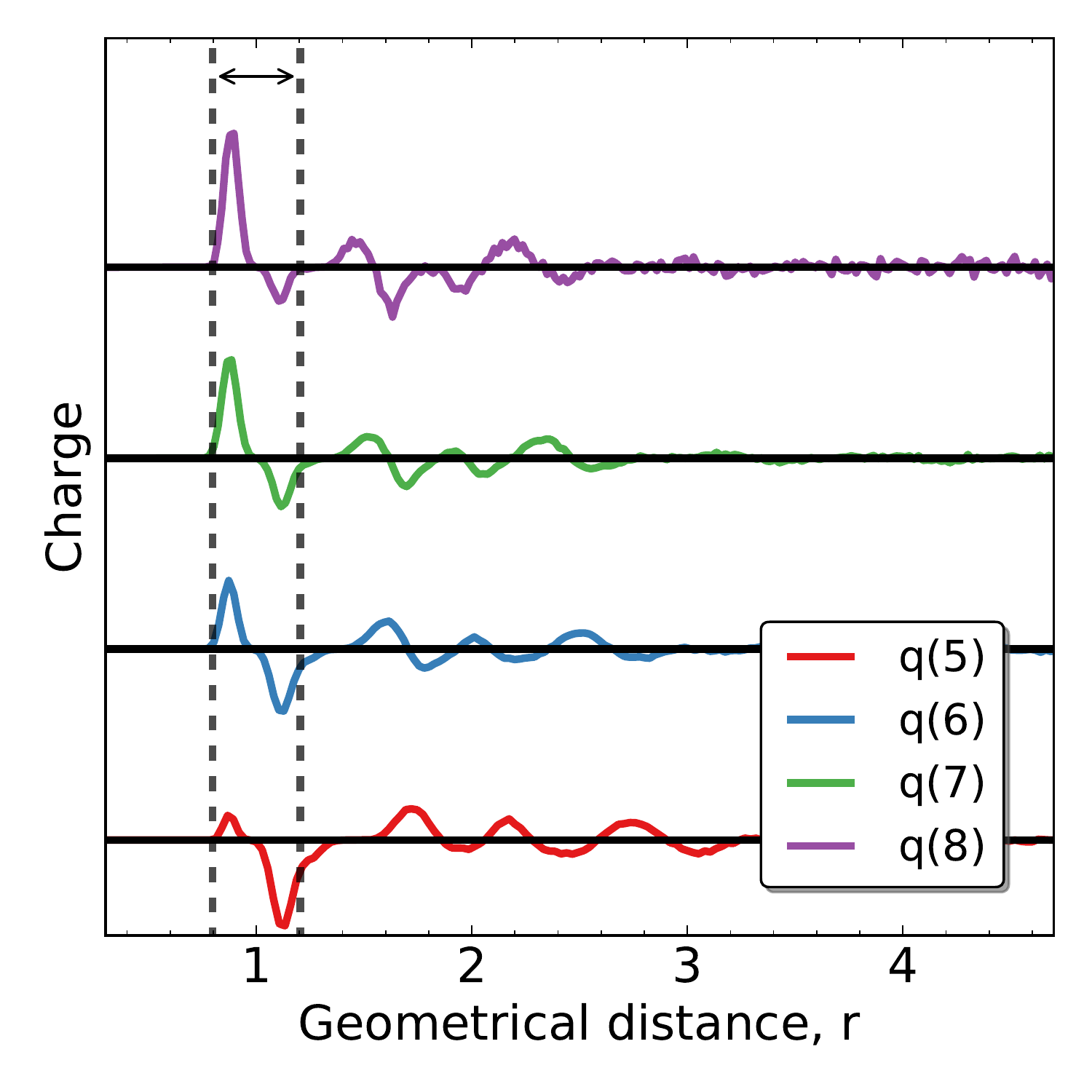}
  \caption{
  \label{fig:chargegeo}
  The topological charge $q_r(n)$ per shell is plotted against the geometrical
  distance $r$. The shell charge approaches zero for distances about three
  rings away.  This figure should be compared to Fig. \ref{fig:charge}. The two
  dashed lines represent the geometrical distance corresponding to the minimum
  and maximum values for the first shell with $t=1$. Curves are offset for clarity
  where horizontal solid lines show the expected asymptotic values of zero.}
\end{figure}

We would like to extend Aboav-Weaire law to longer distances to study
correlation of a ring with the shells around it. The above form can be
used to propose a generalized Aboav-Weaire law as:
\begin{equation}\label{eq:nmn}
  n m_t(n) =  \braket{n}^2 + \mu_t + \braket{n} (1-\alpha_t)(n-\braket{n}),
\end{equation}
where for $t=1$ we recover Eq. \ref{eq:awl} with $\mu_{1}=\mu$.
A similar argument presented to derive Eq. \ref{eq:asymcharge} can be used
to find an asymptotic value for $m_t(n)$. At sufficiently long distances, the
ring distribution in the shells is independent of the size of the central
ring and $\braket{M_t} \approx \braket{m_t K_t} =
\braket{m_t} \braket{K_t}$, therefore for $t>\xi$:
\begin{equation}\label{eq:asvm}
\braket{m_t} = \frac{\braket{M_t}}{\braket{K_t}} =
\frac{\braket{nK_t}}{\braket{K_t}} = 6 - \frac{\braket{q_{\infty}}}{\braket{K_t}}.
\end{equation}
While we expect $\alpha_{\infty}=0$ but we showed, $\braket{q_{\infty}}<0$,
so the asymptotic value of $m_{\infty}(n)$ is larger than the bulk value $6$.
For this reason, $m_t(n)$ approaches 6 as $t^{-1}$ (since $K_t(n) \sim t$)
which is sometimes interpreted as a long-range correlation \cite{
wang2012generalization,wang2014long}. However this should be regarded as an
artifact because the shells are defined in such a way (topologically) which
results(unfortunately) in the topological charge never going to zero, even at
very large distances, and in fact approaching a constant as shown here. This
is due to the non-circular nature of the shells, and can be avoided if the
shells are chosen in such a way as to make them more nearly circular.
Unfortunately this is not possible with a purely topological definition, and
so we are forced to adopt a \textit{geometrical definition} for the ring-shell
correlations.

Figure \ref{fig:dis} shows the difference between topological and geometrical
distance. Despite the fact that shells found by topological distance are roughly
circular, it is not possible to find a single circle which contains all the
rings in the shell, therefore ring distributions etc. are different in the two cases.

The \textit{geometrical } distance $r$ between two rings is defined as the Euclidean
distance between their centroids. Therefore, instead of using the discrete integer
 distance $t$, the quantities $q$ and $m$ are written as a function of a
 continuous distance $r$:
\begin{align}\label{}
  q_r(n) &= 6 K_r(n) - M_r(n), \\
  n m_r(n) &=  \braket{n}^2 + \mu_r + \braket{n} (1-\alpha_r)(n-\braket{n}).
\end{align}

\begin{figure}[t]
  \centering
  \includegraphics[width=5.8cm,height=5.8cm]{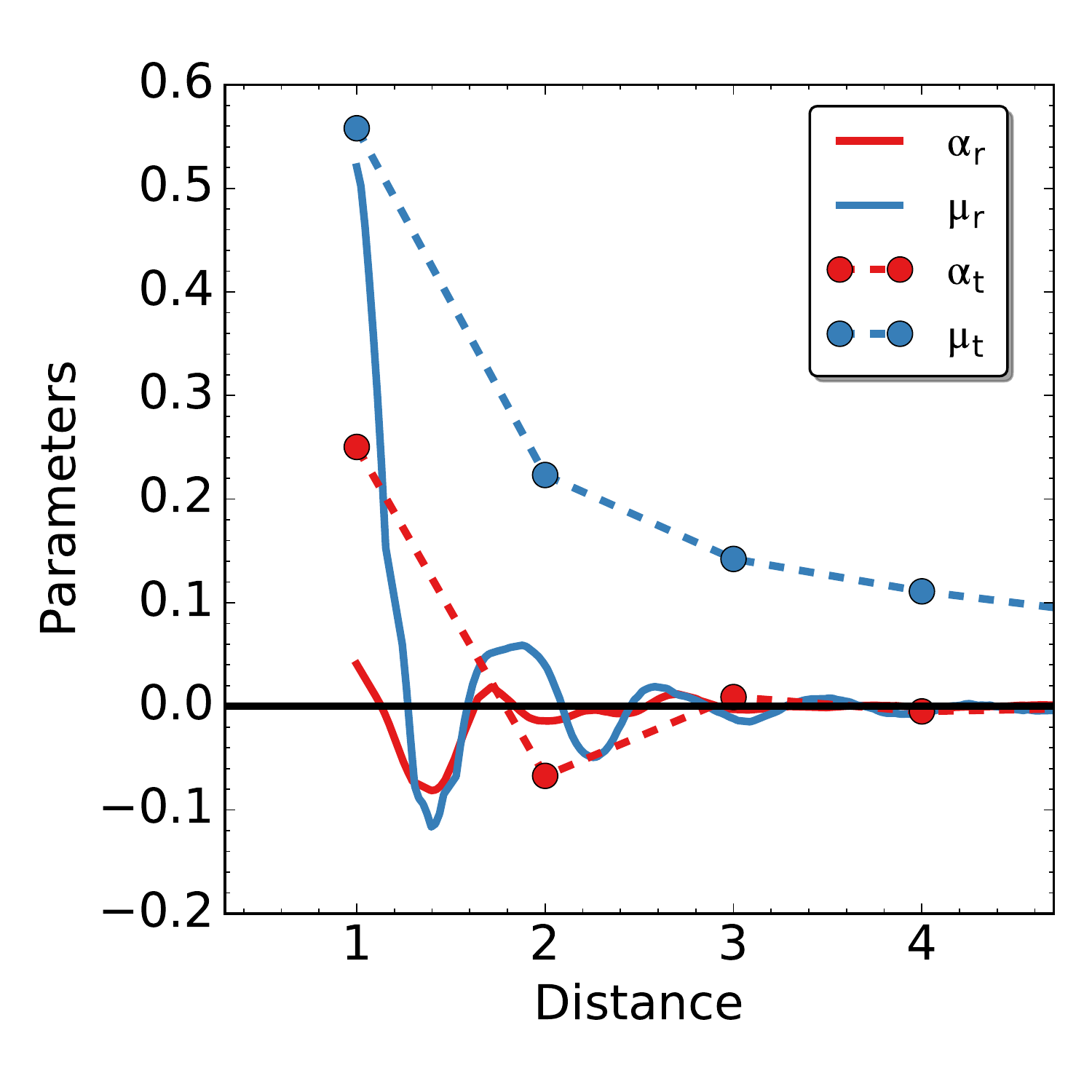}
  \caption{\label{fig:aboav}
  Plot of two coefficients in the generalized Aboav-Weaire law,
  $\alpha_r$ and $\mu_r$ with their topological counterparts,
  $\alpha_t$ and $\mu_t$. Geometrical definitions show that
  correlation quickly decays to zero while pseudo-correlations
  in the topological case last over a long-range for $\mu_t$.  The geometrical
  distances are chosen so the geometrical and topological distances agree
  for the first shell.}
\end{figure}

Since $r$ is continuous, a binning procedure is used to compare with the previous
results using topological distance. Small bins are used with a windowing
procedure where the width of the window mimics unity in topological distance.
Results for the charge are shown in Fig. \ref{fig:chargegeo}. It is evident
that correlations last about $3$ shells and are quite short-ranged with the
charge going to $0$ over the same range, as expected. Therefore this definition
of a shell using geometrical distance is more useful. Because of the different
size of the rings, e.g., distance between a $5-6$ pair is greater than a $7-8$
pair so a range of geometrical distances corresponds to a single topological
distance. To compare the two distances, we rescale the geometrical distance
by the average distance between adjacent rings, which is defined to be unity.
 Fig. \ref{fig:chargegeo}
shows this for the first neighbors with two dashed lines. Within this
window, all four curves show a common trend: a maximum followed by a
minimum. The former corresponds to $5-$rings (positively charge) and
the latter to $7-$ and $8-$rings (negatively charged). The point in the middle
corresponds to neutral $6-$rings. The horizontal axis is normalized such
that these three points line up for all curves. According to Aboav-Weaire
law, smaller rings surround a larger ring; the pronounced minimum
of $q_r(5)$ due to $7-$ and $8-$rings and the pronounced maximum of $q_r(7)$
and $q_r(8)$ due to $5-$rings admit this law. In the case of $q_r(6)$,
minimum and maximum have the same amplitude due to uniform distribution
of the rings around hexagons hence their weak correlations with other
rings.

It is also constructive to look at Aboav-Weaire law using geometrical
distance. In this case, we expect that both $\alpha_{r}$ and $\mu_{r}$ decay
rapidly to zero in accordance with the absence of correlations for large
$r$. This is confirmed in Fig. \ref{fig:aboav} which clearly for distances
larger than $3$, the mean ring size is essentially exactly $6$.
This confirms our assertion that ring correlations in glassy networks are
either short-range or medium-range and using geometrical distance in the
calculations of topological charge and mean ring size resolves the issue
of excess topological charge in the shells found by topological distance
which is shown by the long-tail of $\mu_2$ in Fig. \ref{fig:aboav}.

\begin{figure}[t]
  \centering
  \includegraphics[width=5.8cm,height=5.8cm]{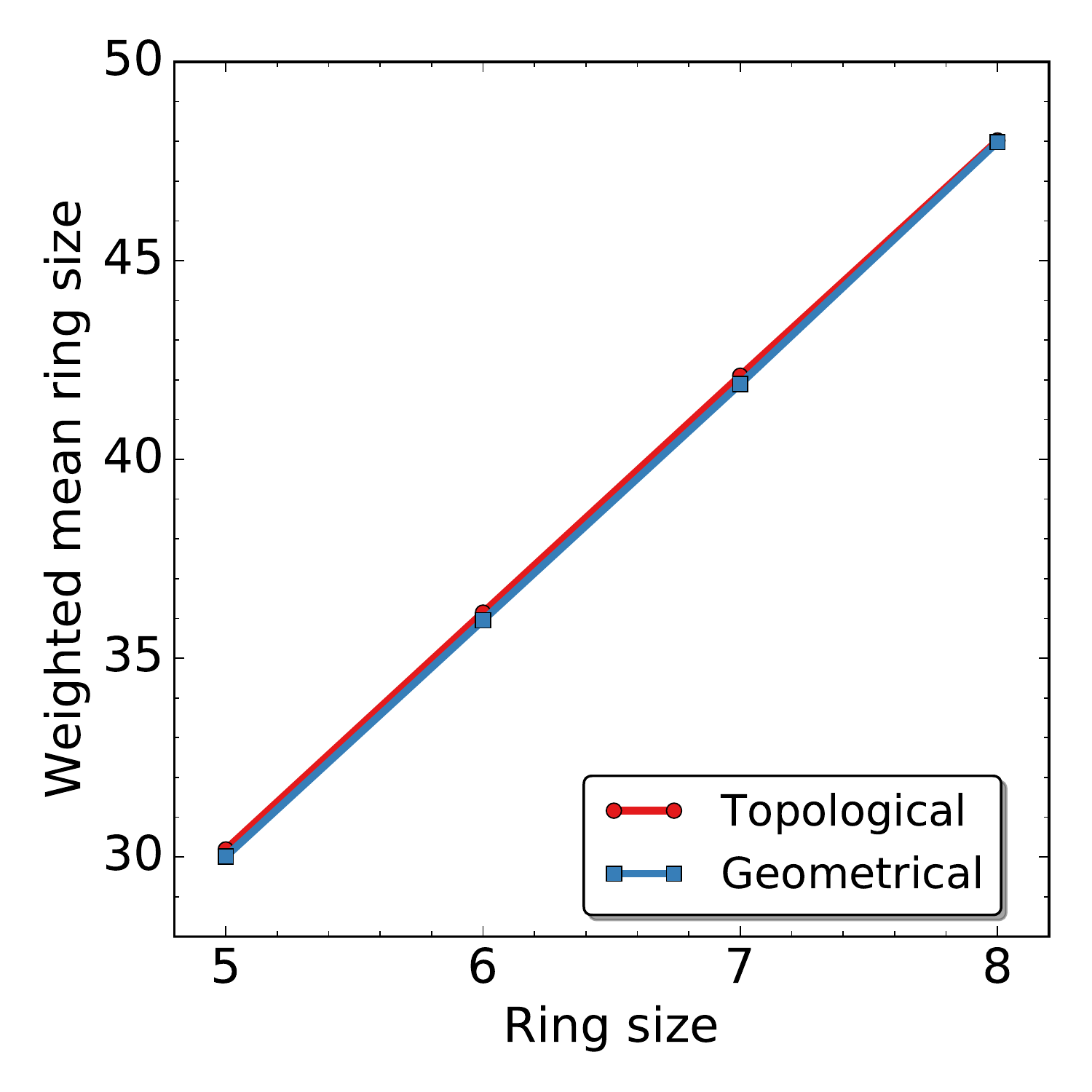}
  \caption{
  \label{fig:nmn}
  Plot of weighted mean ring size $nm(n)$ versus ring size $n$ for
  the third neighbors using both geometrical and topological distance. This plot
  shows that the mean ring size for all shells follows the generalized
  Aboav-Weaire law (Eq. \ref{eq:nmn}). Note that the topological definition
  leads to a slightly larger mean ring size.}
\end{figure}

Fig. \ref{fig:nmn} shows linearity of the generalized Aboav-Weaire
law for the third neighbors. The plot shows that $nm(n)$ is indeed a linear
function of $n$ but because of pseudo-correlations, the average ring size using
topological distance is slightly larger than expected for geometrical
distance, where the mean ring size is 6 for three-fold coordinated
networks.

Although the topological charge and Aboav-Weaire law are useful tools to quantify
correlations, they only measure correlations between a ring and shells.
The ring-ring correlation function is perhaps a better measure of
correlations especially since, as it was shown, definition of shells using
the topological distance do introduce some artifacts such as excess charge.

To find out the correlation between two single rings, we need to derive an
expression for the probability $p_t(n,n')$ of finding a pair of $n, n'$ rings
with distance $t$. For a given $n-$ring, the number of $n'-$rings
at distance $t$ is $N_t(n,n')$ while on average a typical shell has
$\braket{K_t}$ rings. Therefore the probability of having a pair of rings
is \cite{szeto1998topological}:
\begin{equation}\label{probability}
 p_t(n,n') = \frac{p(n)N_t(n,n')}{\braket{K_t}}.
\end{equation}

This equation is important as it relates ring distributions of the shell
structure to of the network (For $t=1$, this equation reduces to the
correlation function defined in Ref. \cite{le1993correlations}).
If the rings were independent, this probability is simply product of
the individual probabilities but we showed the ring distribution
of a shell is different from the bulk and rings are topologically
dependent even for large $t$. This motivates us to define the
probability of having an $n-$ring at shell $t$ (independent of the
central ring) as:
\begin{equation}\label{shellprob}
 p_t(n)=\sum_{n'} p_t(n',n) = p(n)\frac{K_t(n)}{\braket{K_t}},
\end{equation}
which can be derived using Eqs. \ref{eq:shellsize} and \ref{eq:symfun}.
The probability of having $n-$ring is proportional to the average shell
size around $n-$fold rings and the ensemble averaged shell size.
We define correlation function between two $n$ and $n'$ sided rings as:
\begin{equation}\label{eq:rrc}
 C_t(n,n') = p_t(n,n') - p_t(n)p_t(n')
\end{equation}

\begin{figure*}[t]
  \centering
  \includegraphics[width=11cm,height=11cm]{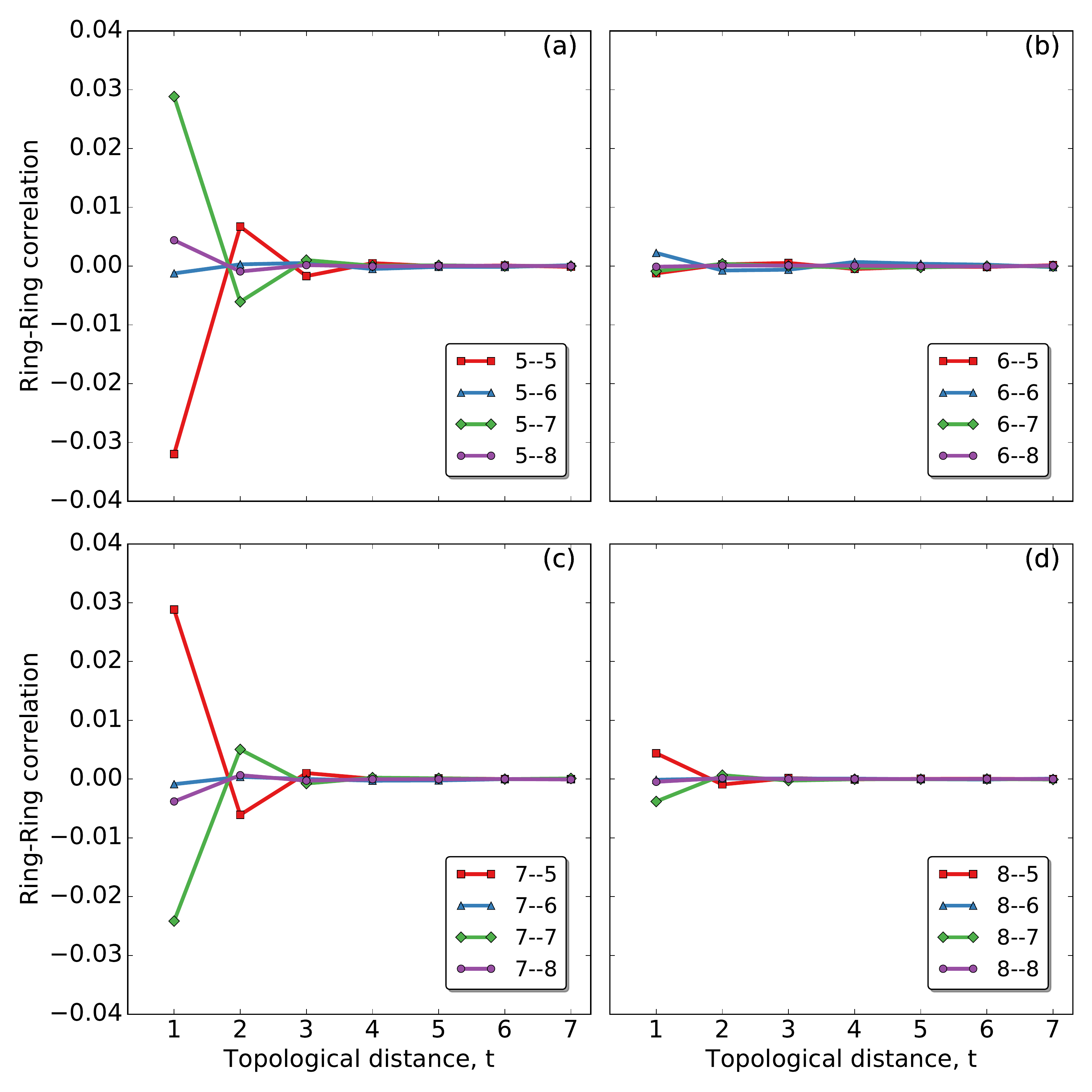}
  \caption{\label{ringringcor}
  Ring-ring correlation $C_t(n,n')$ versus topological distance $t$.
  The correlations are short or medium range depending on the size of the
  interacting rings. Although hexagons are weakly correlated with their
  neighbor rings, other rings show a high degree of correlations up to
  three rings away. Very similar results are obtained using geometrical
  distances. Note that correlations are symmetric so that $5-6$ is
  the same as $6-5$ etc. where panel (a) is for five-fold rings, panel (b)
  six-fold rings, panel (c) seven-fold rings, and panel (d) eight-fold rings.}
\end{figure*}

Figure \ref{ringringcor} shows the results for the above correlation function.
This clearly shows the medium-range order of the rings except for hexagons
where correlations are weak and short-range. In contrast with the results
in Ref. \cite{szeto1998topological}, hexagon-hexagon is short-range
and only non-zero for adjacent cells ($t=1$) which is a signature
of microcrystal regions in the network (see Fig. \ref{fig:randnet}). If
we had used $p(n)p(n')$ instead of $p_t(n)p_t(n')$, ring-ring correlation
shows a long-range behavior due to topological effect
\cite{oguey2011long,miri2007topological} but Eq. \ref{shellprob}
correctly captures the nature of correlations in the random network.

\section{Discussion and Conclusion}

We have shown that correlations between rings in glassy networks can be treated
best if geometrical rather than topological distances between rings are used.
Using topological distances, which would be preferable, unfortunately leads to
spurious long range correlations as the topological charge for each shell around
a central ring does not approach zero at large distances, due to the non-circular
nature of the shells.  These issues are absent if the geometrical distances between
the centers of rings are used. We find in this case that correlations only extend
out to about third neighbor rings, and can be described by a generalized
Aboav-Weaire law. These studies have been done on a very large computer-generated
network with periodic boundary conditions~\cite{wooten1985computer,stone1986theoretical}.
Experimental samples of bilayer of
vitreous silica are currently too small to allow for the study of longer range
correlations, but the main conclusion of the paper that geometrical rather
than topological distances should be used is expected to hold. Future studies
comparing experimental and computer-generated networks (both three-coordinated
with similar ring distributions) should help explain why different values
of $\alpha$ are obtained in these two cases~\cite{kumar2014}.

\begin{acknowledgments}
 We should thank Avishek Kumar for providing the computer-generated networks,
 and David Sherrington and Mark Wilson for useful ongoing discussions. MS was
 partially supported by the Arizona State University Graduate and Professional
 Student Association's JumpStart Grant Program. MS was aided in this work by
 the training and other support offered by the Software Carpentry project.
 Support through NSF grant \# DMS 1564468 is gratefully acknowledged.
\end{acknowledgments}

\bibliography{reference}

\end{document}